\documentclass[useAMS,usenatbib]{mn2e}

\usepackage{graphicx}
\usepackage{aas_macros}
\usepackage{color}
\usepackage{enumerate}
\usepackage{url}
\usepackage{bm}

\title[FiBY: GRBs at $z>5$]{The First Billion Years project: gamma-ray bursts at $z>5$}
\author[Elliott et al.]{J. Elliott$^{1}$\thanks{E-mail: jonnyelliott@mpe.mpg.de}, S. Khochfar$^{2,1}$, J. Greiner$^{1}$, C. Dalla Vecchia$^{3,4}$\\
  $^{1}$ Max-Planck-Institut f{\"u}r extraterrestrische Physik, Giessenbachstra{\ss}e 1, D-85748, Garching, Germany.\\
  $^{2}$ Institute for Astronomy, University of Edinburgh, Royal Observatory, Edinburgh EH9 3HJ, UK.\\
  $^{3}$ Instituto de Astrof{\'i}sica de Canarias, C/ V{\'i}a L{\'a}ctea s/n, 38205 La Laguna, Tenerife, Spain. \\
  $^{4}$ Departamento de Astrof{\'s}ica, Universidad de La Laguna, Av.~del
   Astrof{\'i}sico Franciso S{\'a}nchez s/n, 38206 La Laguna, Tenerife, Spain. \\
}

\begin{document}

\date{Accepted 2014 November 13. Received 2014 October 22; in original form 2014 August 11}

\pagerange{\pageref{firstpage}--\pageref{lastpage}} \pubyear{2014}

\maketitle

\label{firstpage}

\begin{abstract}
Long gamma-ray burst's (LGRB's) association to the death of massive stars suggest they could be used to probe the cosmic star formation history (CSFH) with high accuracy, due to their high luminosities. We utilise cosmological simulations from the First Billion Years project to investigate the biases between the CSFH and the LGRB rate at $z>5$, assuming various different models and constraints on the progenitors of LGRBs. We populate LGRBs using a selection based on environmental properties and demonstrate that the LGRB rate should trace the CSFH to high redshifts. The measured LGRB rate suggests that LGRBs have opening angles of $\theta_{\rm jet}=0.1\degr$, although the degeneracy with the progenitor model cannot rule out an underlying bias. We demonstrate that proxies that relate the LGRB rate with global LGRB host properties do not reflect the underlying LGRB environment, and are in fact a result of the host galaxy's spatial properties, such that LGRBs can exist in galaxies of solar metallicity. However, we find a class of host galaxies that have low stellar mass and are metal-rich, and that their metallicity dispersions would not allow low-metallicity environments. Detection of hosts with this set of properties would directly reflect the progenitor's environment. We predict that 10\% of LGRBs per year are associated with this set of galaxies that would have forbidden line emission that could be detected by instruments on the {\it James Webb Space Telescope}. Such a discovery would place strong constraints on the collapsar model and suggest other avenues to be investigated.
\end{abstract}



\section{Introduction}
\label{sec:Introduction}

Reaching luminosities as high as $10^{54}\, \rm erg\, s^{-1}$ makes long gamma-ray bursts (LGRBs) the brightest objects in the Universe during their emission. With such an unexcelled brightness, LGRBs are the perfect tool for investigating the high redshift Universe once their physical origin is well understood.

In the collapsar model~\citep{Woosley93a} LGRBs are believed to be the end result of a rapidly rotating, massive star that undergoes gravitational collapse. If the newly created compact object (black hole/neutron star) retains angular momentum of the order of $3<j/10^{16}\,\rm cm^{2}\, s^{-1}<20$~\citep[][]{MacFadyen99a}, a disc is formed, which it then accretes from. Relativistic fireball shells are then released in to jetted outflows as a result of the geometry of the disc or magnetic effects~\citep[e.g.,][]{Blandford77a,Narayan92a,MacFadyen99a}. 
Depending on the variability of the accretion rate from the disc, shells are released with different Lorentz factors.
 These shells can then cross as they catch up to one another and a relativistic \emph{internal} shock occurs~\citep{Sari98a}. 
 At the shock front the electron population is Fermi accelerated~\citep{Fermi49a} and then cools via synchrotron emission, which is seen as observer frame $\gamma$-rays.

Many LGRBs have been associated to supernovae~\citep{Galama99a,Stanek03a,Hjorth03a,Matheson03a}, all of which lacked any hydrogen or helium lines and so were of the type-Ib/c~\citep[e.g.,][]{Filippenko97a}. 
In combination with the fact that any outer hydrogen layer of the star would stop the LGRB escaping~\citep{MacFadyen99a}, means that the hydrogen envelope of the progenitor has to be lost before it undergoes gravitational collapse. 
This has lead to the favoured LGRB progenitor, a Wolf-Rayet-like star. 
However, these massive stars have extremely large radiative-line driven winds~\citep[e.g.,][]{Vink05a} that lead to angular momentum loss, stopping the formation of a LGRB~\citep{Hirschi05a,Yoon05a,Woosley06a}. As winds are highly dependent on metallicities, such that low-metal envelopes have weaker winds~\citep[$\dot{M}\propto Z^{x}$ for $x>0$, e.g.,][]{Puls08a}, it has been proposed that LGRBs can only form in environments with low-metallicities~\citep[$0.3\,\rm Z_{\odot}$, e.g.,][]{Hirschi05a} via the single progenitor model.

The association of LGRBs with the death of massive stars~\citep{Galama99a,Stanek03a,Hjorth03a,Matheson03a} facilitates their use as star formation tracers and their naturally luminous nature would complement or even surpass conventional methods at high redshift. However, before this connection can be used routinely and robustly, the way in which they trace one another must be known accurately.

This has lead many authors to investigate the differences, if any, between the LGRB rate and the cosmic star formation history~\citep[CSFH; e.g.,][]{Wijers98a,Bromm06a,Daigne06a,Li08a,Butler10a,Wanderman10a, Elliott12a}. Early studies showed an overabundance of LGRBs at higher redshifts than that inferred from the CSFH~\citep[e.g.,][]{Daigne06a}. This underlying excess has been explained for many different reasons ranging from observational biases~\citep{Coward08a, Elliott12a} to redshift-dependent initial mass functions~\citep[IMF;][]{Wang11a}. However, the most favoured explanation is a metallicity dependence that is driven by (i) the requirement of the single progenitor collapsar model, and (ii) LGRB host galaxy observations~\citep[see, e.g.,][]{LeFloch03a, Savaglio09a, Svensson10a, Mannucci11a}.

Given the difficulty of probing the progenitor itself, empirical proxies have been used to model the inherent bias of the LGRB rate. The first results obtained on host galaxy samples showed that they were primarily blue with low-mass, and low-metallicity~\citep[e.g.,][]{Fruchter99a, LeFloch03a, Berger03a, Christensen04a, Tanvir04a, Savaglio09a} and suggested that the metallicity of the host could be a proxy for the progenitor. Such proxies exerted a strict metallicity-cut~\citep[e.g.,][]{Langer06a,Salvaterra07a,Butler10a,Elliott12a}, such that only host galaxies below a given value could host a LGRB, which lead to predictions that the CSFH flattens at higher redshifts~\citep[e.g.,][]{Kistler09a,Salvaterra12a}. However, these host galaxy studies were optically biased, which when accounted for, systematically more massive galaxies were found~\citep[e.g.,][]{Kruehler11a,Rossi12a}. Despite this increase in more massive galaxies, there still existed a deficit of massive, metal-rich host galaxies~\citep[e.g.,][]{Graham12a,Perley13a}, which lead to the strict metallicity cut-off being increased to $Z<0.5\,\rm Z_{\odot}$.

There is now mounting evidence of LGRBs existing in solar/super-solar metallicity galaxies, which makes it difficult to reconcile with the strict metallicity-cut off~\citep[e.g.,][]{Levesque10c,Savaglio12a,Kruehler12b,Elliott13a}. To this end, recent studies have attempted to represent the host galaxy metallicity bias as a probability distribution, whereby, the more metal-rich environments contribute less to the quantity of LGRBs~\citep{Trenti13a}.

Given the lack of robust host galaxy integrated metallicity measurements above $z\sim1$ and sight-line measurements of metallicity at the location of the LGRB, make it difficult to investigate any dependences that may exist and near-impossible to identify any evolution of the progenitor model. In addition, the proxies that describe the dependence do not necessarily draw any direct conclusions to the progenitor.

In contrast, the ease of use of cosmological simulations has improved over recent years that it is becoming more common to use them to make predictions about LGRBs~\citep[e.g.,][]{Campisi11a}, their environment~\citep[e.g.,][]{Courty04a,Nuza07a,Pontzen09a,Chisar10a,Artale11a}, and the influence that different effects in the Universe have on them~\citep[e.g.,][]{Maio12a}. 
Despite their accessibility, the recent use of cosmological simulations have poor mass-resolutions~\citep[e.g.,][]{Campisi11b}, and usually have limited or no chemical treatment during run time, which makes it difficult to investigate the local environments and properties of LGRBs.

The aim of this paper is three-fold: Firstly, within a high-resolution cosmological simulation we self-consistently compare LGRB rates determined by (i) populating LGRBs using a physical model based on stellar collapse models and (ii) analytical proxies utilising the CSFH. Secondly, we compare a selection of analytical proxies, that attempt to connect local LGRB properties to the global properties of host galaxies, to see if they can be used reliably to estimate the LGRB rate. Finally, we investigate the connection, if any, between the progenitor environment and the resulting host galaxy population.

The paper is structured as follows: we first introduce the simulation we use for our study in Sect.~\ref{sec:Simulations}. Section~\ref{sec:Methodology} outlines the methods used to generate a LGRB based on its environment and how we calculate the LGRB rate and CSFH. The results are then displayed in Sect.~\ref{sec:Results}. We then discuss the implications of our findings in Sect.~\ref{sec:Discussion} and conclude in Sect.~\ref{sec:Conclusion}.


\section{Simulation}
\label{sec:Simulations}
The First Billion Years (FiBY) project~(Khochfar et al., in preparation), is a suite of cosmological simulations primarily aimed at investigating the formation and evolution of the first galaxies. The project uses a modified version of the smoothed-particle-hydrodynamics (SPH) code \texttt{GADGET}~\citep{Springel05a}, developed for the Overwhelmingly Large Simulations project~\citep{Schaye10a}. Modifications to the code will be described in detail in Dalla Vecchia et al (in preparation). The reliability of the simulations is to be outlined in future work by Khochfar et al. (in preparation), who will show, for example, that the FiBY project simulations reproduce well the mass function of high-z galaxies and the star formation rate of individual galaxies. The simulations are representative of average density regions, and the dark matter mass functions are in agreement with the expectations of the~\citet{Sheth99a} mass function. We briefly show comparisons of the cosmic star formation history from the simulations to observations in Sect.~\ref{sec:Results:SimulationCSFH}. In the following we give a summary of the physics included in the simulation.

The simulation includes: (i) star formation based on the pressure-law in~\citealt{Schaye08a} (ii) line cooling in photo-ionisation equilibrium for a total of eleven elements (H, He, C, N, O, Ne, Mg, Si, S, Ca, and Fe) following~\citealt{Wiersma09b}, (iii) metal enrichment from type-I/II and pair-instability supernovae and stellar winds~\citep{Tornatore07a, Wiersma09a}, (iv) thermal feedback from supernovae~\citep{DallaVecchia12a}, (v) a full non-equilibrium primordial composition chemical network and molecular cooling for the H$_{2}$ and HD molecules~\citep{Abel97a, Galli98a, Yoshida06a, Maio07a}, (vi) formation of Population-III stars and chemical enrichment from them~\citep{Heger02a, Heger10a}, and (vii) reionisation feedback modelled with a uniform UV background~\citep{Haardt01a} that sets in at a redshift of 12 and reaches photo-ionisation equilibrium at redshift 9. This implementation of reionisation is consistent with the ionising emissivity history determined from the galaxies in the simulation box~\citep{Paardekooper13a}.

Star formation is based on a pressure law, and only takes place above a given density threshold $n_{\rm H}=10\, \rm cm^{-3}$. Above this density threshold we assume a polytropic equation of state for the gas, $P\propto\rho^{\gamma}$, normalised to $T=1000\, \rm K$ at the threshold. We set $\gamma=\frac{4}{3}$ to ensure that the Jeans length is resolved by the same number of SPH smoothing lengths for any $\rho > n_{\rm H}$, thus stopping artificial fragmentation that would yield higher rates of star formation~\citep{Schaye08a}. The population type of stars (II or III) is defined by a strict metallicity limit, such that stars that form in gas with metallicities of $Z<Z_{\rm crit}$ are Population-III stars with a Salpeter IMF~\citep{Salpeter55a} going from $21\, \rm M_{\odot}$ to $500\, \rm M_{\odot}$~\citep{Bromm04b,Karlsson08a}, and for $Z\geqslant Z_{\rm crit}$ are Population-II stars with a Chabrier IMF~\citep{Chabrier03a} going from $0.1$ to $100\, \rm M_{\odot}$, assuming $Z_{\rm crit}=10^{-4}$ ($\rm Z_{\odot}=0.02$). The exact transition from one stellar type to another is still currently not known, but the relative changes to the star formation are small for changes in $Z_{\rm crit}$ in comparison to other uncertain parameters such as the Population-III IMF mass range~\citep{Maio10a}.

Thermal feedback from the supernovae is carried out when the lifetime of the lowest stellar mass going into a supernova is reached~\citep{DallaVecchia12a}. For Population-II stars in the mass range $\left[8,100\right]\, \rm M_{\odot}$ they end their lives as type-II supernovae and release an energy of $\epsilon=10^{51}\, \rm erg$ per supernova assuming a Chabrier IMF. For Population-III stars the same treatment as Population-II stars is applied for the mass range of $\left[21,100 \right]\, \rm M_{\odot}$, but instead a Salpeter IMF is adopted. However, for the mass range of $\left[140,260\right]\, \rm M_{\odot}$ Population-III stars are believed to end their lives as the more energetic pair-instability supernovae for which an energy of $\epsilon=3\times10^{52}\, \rm erg$ per supernova is released~\citep{Heger02a}. The energy from thermal feedback is converted into kinetic energy that results in the dispersal of star forming gas, and thus the suppression of star formation. This suppression is effective for a period of time until the radiative cooling becomes dominant, allowing the gas to collapse once again~\citep{DallaVecchia12a}.

Within the simulation, the three types of metal pollution are stellar winds, asymptotic giant branch stars, and type-Ia/II supernovae. AGB stars of masses $0.8-8\, \rm M_{\odot}$ contribute primarily carbon and nitrogen to the intergalactic medium, oxygen and $\alpha$-elements (e.g., neon, magnesium, and silicon) are produced by the supernovae type-II, and then iron from type-Ia supernovae~\citep{Wiersma09a}. The quantity of metal enrichment, at a given time-step of the simulation, is determined by folding the stellar IMF (for the specific population) with an inverse stellar life-times function determined from stellar evolutionary track models~\citep[][and references therein]{Wiersma09a}. The mass of metal enrichment from supernovae-Ia is calculated from the observational measured supernovae-Ia rate. The quantity of each metal is then determined from tabulated yields for eleven different elements. Within the gas, each of the elements are allowed to radiatively cool utilising the prescription outlined in~\citealt{Wiersma09b}. The metal enrichment history from stellar evolutionary tracks are consistent for a range in metallicities, but can change drastically depending on the underlying IMF chosen~\citep[see,][for more details]{Wiersma09a}.

\subsection{Simulation runs}
The FiBY project consists of a number of simulations with different resolution. In this study we will use the FiBY\_S and the FiBY\_L, which both consist of $648^3$ SPH and dark matter (DM) particles in co-moving volumes of $4^3$ and $16^3$ Mpc$^3$, respectively. We chose these two simulations to bracket the range in resolution and volume that will allow us to probe galaxies with stellar masses of $10^{3}-10^{10}\, \rm M_{\odot}$ and at the same time investigate resolution effects. The mass per SPH particle $m_{\rm sph}$ is $\sim$ 1254  and $\sim$ 80225 $\rm M_{\odot}$, and the mass per DM particle $m_{\rm DM}$ is $\sim$ 6162 and $\sim$ 394366 $\rm M_{\odot}$ for the individual simulations.
 
Within the simulation we cannot resolve individual stars, but in fact can resolve stellar populations with a stellar mass resolution of $m_{\rm sph}$. We refer to these systems as star particles throughout the text. We utilise Friend-of-Friend~\citep{Davis85a} and SUBFIND~\citep{Springel01a} algorithms to define haloes, which require the minimum number of particles (dark matter, gas, and stars) to be 100. Galaxies containing 100 star particles are sufficient to obtain robust values of global galaxy properties~\citep[see, e.g.,][]{Schaye14a}. For example, in FiBY\_L, 100 star particles corresponds to $\sim10^{6.9}\, \rm M_{\odot}$ solar masses, sufficient to resolve metallicities of $12+\log(\rm O/H)\sim 9$.

The simulation adopts the cosmological parameters consistent with the Wilkinson Microwave Anisotropy Probe~\citep[WMAP;][]{Komatsu09a}, i.e., $\Omega_{\rm m}=0.265$, $\Omega_{\rm b}=0.0448$, $\Omega_{\Lambda}=0.735$, $H_{0}=71\, \rm km s^{-1} Mpc^{-1}$, and $\sigma_{8}=0.81$. It was started at a redshift of $z=127$ and finished at a redshift of $z=6$ for the FiBY\_S and at z$=5$ for FiBY\_L.

\subsection{Cosmic Star Formation History}
\label{sec:Results:SimulationCSFH}

\begin{figure}
  \begin{center}
    \includegraphics[width=9.5cm]{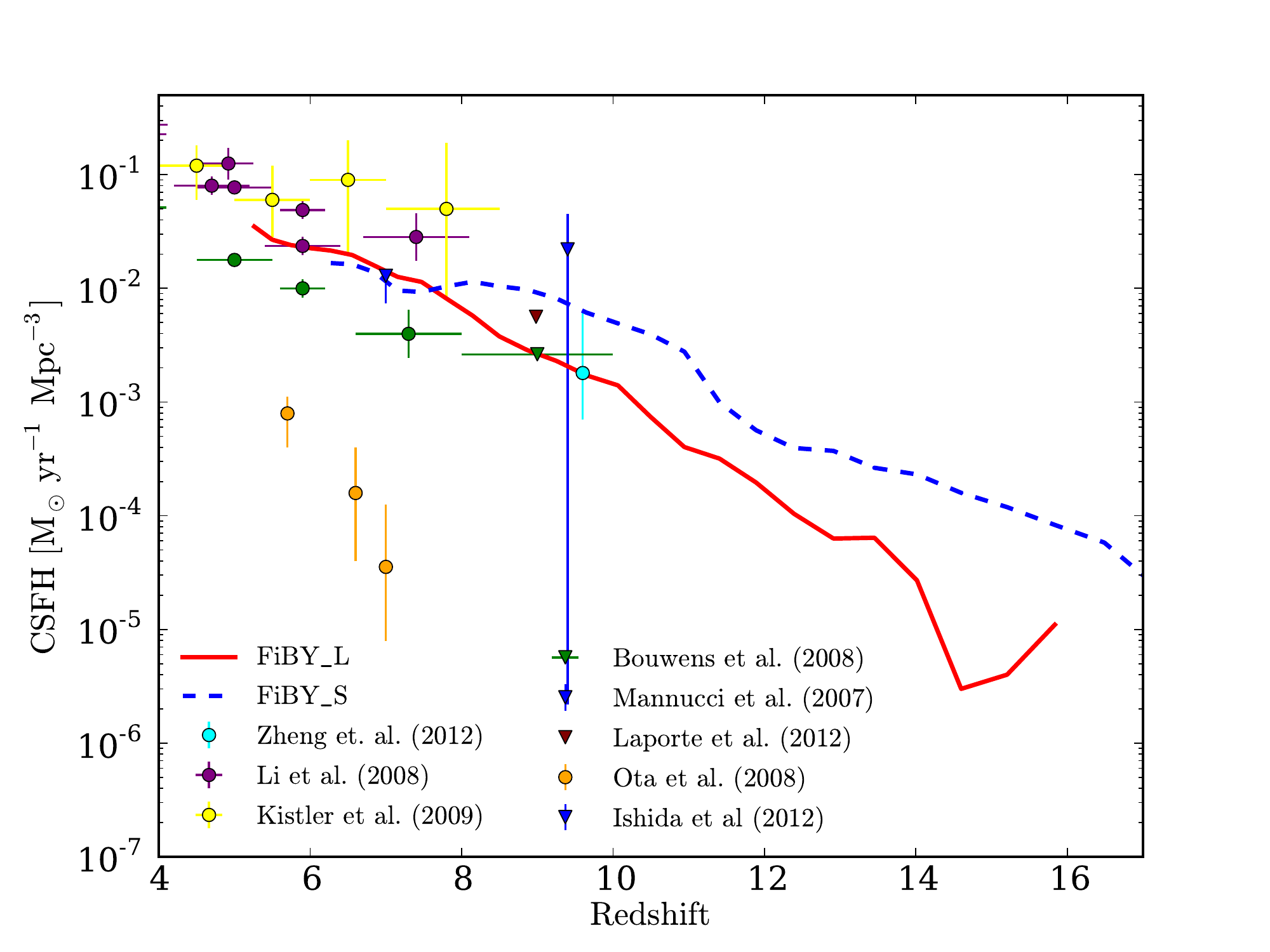}
    \caption{Cosmic star formation history of the small $\left(4\rm Mpc\right)^{3}$ box (FiBY\_S, blue-dashed line) and large $\left(16\rm Mpc\right)^{3}$ (FiBY\_L, red line). Each of the data points are CSFH measurements obtained from the literature and are in good agreement with the simulation. Note that the FiBY\_S simulation has a factor 8 better resolution than the FiBY\_L, which explains the earlier onset of star formation and its higher values at $z>10$.}
    \label{fig:pfb2_csfh}
  \end{center}
\end{figure}

The CSFH of the FiBY simulation is shown in Fig.~\ref{fig:pfb2_csfh}. We compare it to several values obtained from the literature, determined from star-forming galaxies~\citep{Mannucci07a,Bouwens08a,Li08a}, Lyman-break galaxies~\citep{Laporte12a, Zheng12a}, Lyman-$\alpha$ emitters~\citep{Ota08a}, and LGRBs~\citep{Kistler09a,Ishida11a} that reach a redshift of $z\sim10$. Both the CSFH of the FiBY simulation and the measured distribution are seen to be consistent throughout redshift.

\section{Model \& Methodology}
\label{sec:Methodology}
In the following two sub-sections we describe how we implement LGRBs into our simulation using (i) a model based on the physics of progenitor stars and (ii) a series of models that rely on proxies derived from host galaxy properties.

\subsection{The Progenitor Model}
In this paper we address only LGRBs originating from Population-II stars and defer the reader elsewhere for Population-III LGRBs~\citep[see, e.g.,][]{Nagakura12a}. The latter, however, will only play a minor role in the overall LGRB population between $5<z<20$ given the dominance of Population-II star formation at these redshifts.

\subsubsection{Initial Mass Function} 
\label{sec:ModelandMethodology:subsec:InitialMassFunction}
The IMF quantifies how many stars per stellar mass $m$ are created during star formation (for a complete overview see~\citealt{Bastian10a}) and is quantified in the following way

\begin{equation}
  \psi\left(m\right) \equiv \frac{\mathrm{d}n}{\mathrm{d}m} \propto m^{-\alpha},
\end{equation}

\noindent where $\alpha$ depends on the adopted IMF. The IMF for Population-II stars is usually directly measured for nearby star clusters and locally resolved stellar populations~\citep{Bastian10a}, and occasionally indirectly via dynamical modelling~\citep[][]{Cappelari12a}. 
At higher redshifts, it is only inferred from integrated galaxy properties or other methods~\citep[see, e.g.,][]{Dave08a}. There is still no current consensus on what shapes the IMF and thus how it varies with redshift or environment~\citep[see, e.g.,][]{Elmegreen97a,Padoan02a,Larson05a}. Throughout this work we adopt, for Population-II star particles, a Salpeter~\citep{Salpeter55a} IMF ranging from $0.1-100\,\rm M_{\sun}$ ($\alpha=2.35$) that is constant throughout redshift. We note we have used a different IMF to that of the simulation. As our stochastic sampling of LGRBs is redshift independent changing the IMF would only change the overall normalisation (or absolute value of LGRBs). This, however, has no effect on the results outlined in this paper and any change in the overall rate of LGRBs caused by changing the IMF would be much smaller than the other model parameters involved (we discuss this in more details in Sect.~\ref{sec:Discussion:ModelLimitations}).

We sample for each star particle the IMF using the Monte-Carlo Alias method~\citep{Kronmal79a}. For simplicity we assign the metallicity of the star particle at formation to each of the populated stars, but we note that there could be local metal dispersions within the star particle itself. For example, observations have shown that low-metallicity globular clusters can have little or large metallicity dispersions, which even vary depending on the species considered~\citep[see, e.g.,][and references therein]{Roederer11a}. However, note that high-redshift galaxies are predicted to have high levels of turbulence~\citep{Burkert10a} that would efficiently mix metals and thus support the assumption of a single metallicity per star particle.

Sampling of the IMF continues until all of the star particle's mass is allocated. Given that there is a lower bound at which no BH can result in a LGRB within the stellar explosion model we consider (see Sect.~\ref{sec:ModelandMethodology:subsec:ProgenitorMassMetallicity}), we remove the total mass attributed to these stars to decrease the total processing time, and stop sampling the IMF once the remaining mass drops below the lower mass bound ($M_{*}=29\, \rm M_{\odot}$, for a full list of parameters see Table~\ref{tab:model_parameters}). By sampling the IMF directly, and not just assigning a probability for a star particle to host a LGRB progenitor, we allow a single star particle to host more than one LGRB with progenitor stars of different masses. The latter is particularly important for cases that have low-mass resolution, which would not have the ability to apply a mass-metallicity relation for progenitor stars (see Sect.~\ref{sec:ModelandMethodology:subsec:ProgenitorMassMetallicity}). We note that only two realisations were carried out for each simulation due to computational constraints. The star particles in FiBY\_L contain enough mass that a single realisation reproduces well the IMF at larger masses, that several realisations would not alter our conclusions. The FiBY\_S is more sensitive, due to the smaller masses, but as will be shown in Sect.~\ref{sec:Results}, it mirrors the results obtained from FiBY\_L. Therefore, despite our limited sampling in FiBY\_S, carrying out more realisations would not alter our conclusions.

\subsubsection{Progenitor Mass-Metallicity}
\label{sec:ModelandMethodology:subsec:ProgenitorMassMetallicity}
\begin{figure}
 \begin{center}
  \includegraphics[width=9.5cm]{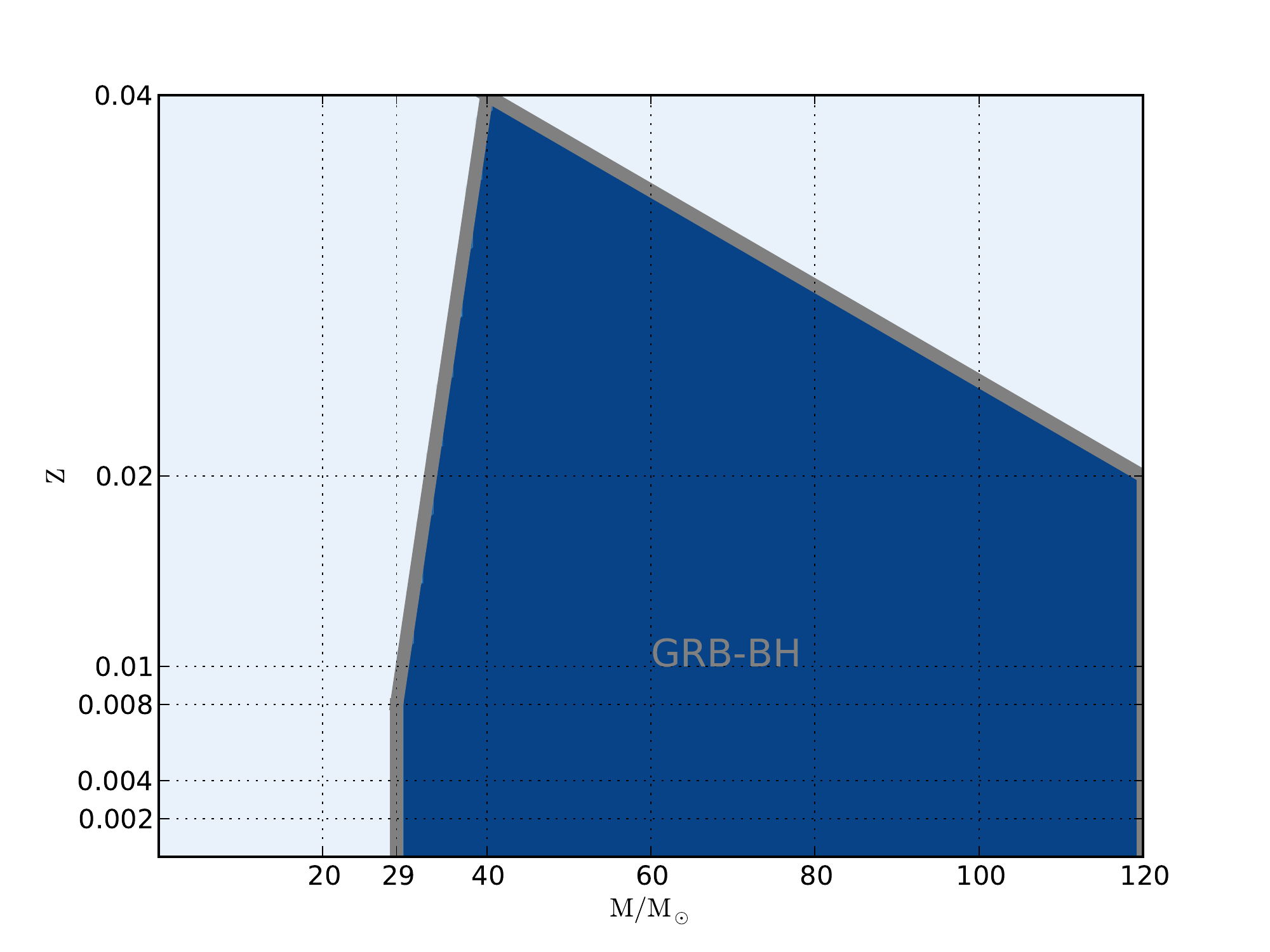}
  \caption{The metallicity plotted against the stellar mass of a grid of Wolf-Rayet progenitor stars taken from~\citet{Georgy09a}. The navy shaded region bounded by grey lines encloses the mass-metallicity region in which a LGRB will occur.}
  \label{fig:mass-metallicity}
 \end{center}
\end{figure}

The collapsar model, which describes how a viable LGRB BH forms, usually requires lower metallicity gas to ensure that angular momentum is retained and a disc is formed~\citep[e.g.,][]{Hirschi05a}. 
Techniques that include this effect when creating LGRBs are often applied to global properties of the host galaxy, via proxies, rather than the LGRB itself~\citep[e.g.,][]{Hao13a}.
 We employ a mass-metallicity relation for LGRB progenitors based on grid simulations of Wolf-Rayet stars~\citep{Georgy09a} that result in LGRB-BHs, which is depicted in Fig.~\ref{fig:mass-metallicity}.
 We consider sampled stars that lie within the LGRB-BH region (navy) as viable LGRB progenitors, whereas stars that lie in the separate part of the phase-space (light blue) regions will explode to another class of compact object that is outlined more thoroughly in~\citet{Georgy09a}.

\subsubsection{Progenitor Age}
\label{sec:ModelandMethodology:subsec:ProgenitorAge}
We have assumed a Wolf-Rayet type progenitor and thus a star that has a short lifetime in comparison to the average stellar population. As a consequence, at any point in our simulation we only consider sampled stars with ages that are old enough to end their lives, but are not from star particles, for which their Wolf-Rayet star population would have already occurred. The upper age limit is also limited so that it is not larger than the time steps between each output of the simulation to avoid double counting. We take the lifetimes of different Wolf-Rayet types from a grid of stellar explosion models by~\citet{Georgy12a}, which give a median of $10^{5.56}\,\rm yr$, with a spread of $\pm2\,\rm dex$. The lower limit is therefore set to $\tau_{\rm lower}=10^{3.56}\,\rm yr$ and the upper limit to our shortest simulation time step, $\tau_{\rm upper}=10^{6.92}\,\rm yr$.

	\begin{table}
	\begin{center}
	  \caption{Parameters used in the progenitor model}
	  \label{tab:model_parameters}
	  \begin{tabular}{l l p{4cm}}
	  \hline
	  Parameter & Value & Description\\
	  \hline
	  IMF index, $\alpha$ & $2.35$ & Power index of the IMF used \\
	  $M_{\rm BH}$ & $29\, \rm M_{\odot}$ & Lowest mass assumed to create a BH/GRB \\
	  $M_{\mathrm{min}}$ & $0.1\, \rm M_{\odot}$ & Lower mass limit of the IMF used \\
	  $M_{\mathrm{max}}$ & $100\, \rm M_{\odot}$ & Upper mass limit of the IMF used \\
	  $\tau_{\rm lower}$ & $10^{3.56}\, \rm yr$ & Lowest age at which a Wolf-Rayet star could die \\
	  $\tau_{\rm upper}$ & $10^{6.92}\, \rm yr$ & Highest age at which a Wolf-Rayet star could die\\
	  \hline
	  \end{tabular}
	
\end{center}
\end{table}

\subsection{The Proxy Model}
\label{sec:ModelandMethodology:subsec:ProxyModel}

\subsubsection{The Connection}
\label{sec:ModelandMethodology:subsec:ProxyModel:subsubsec:TheConnection}
To determine the CSFH of the FiBY simulations, we calculate the total mass of stellar particles for each redshift simulation-output and compute the rate in change of stellar mass to the previous redshift output, $\dot{m}_{*} \equiv (m_{*,n}-m_{*,n-1})/(t(n)-t(n-1))$. This approach was chosen so that the average CSFH could be matched, in redshift, with the properties of the star particles per each simulation output. The final CSFH, $\dot{\rho}_{*}$, is obtained by normalising $\dot{m}_{*}$ to the box size of the simulation.

The inferred association of the death of massive stars with LGRBs allows LGRBs to be used as proxies of massive star formation and, on cosmological scales, the CSFH via the following equation:

\begin{eqnarray}
\label{eqn:csfh_lgrbr}
\frac{\mathrm{d}N\left(z_{1},z_{2}\right)}{\mathrm{d}z} = \frac{1}{z_{2}-z_{1}}\frac{\int_{M_{\mathrm{BH}}}^{M_{\mathrm{max}}} \psi\left(m\right)\,\mathrm{d}m}{\int_{M_\mathrm{min}}^{M_{\mathrm{max}}} m\psi(m)\,\mathrm{d}m}\int_{z_{1}}^{z_{2}} \frac{\dot{\rho_{*}}\left(z\right)}{\left(1+z\right)}\,\mathrm{d}z,
\label{eqn:grb_rate_fiby}
\end{eqnarray}
\noindent where $N\left(z_{1},z_{2}\right)$ is the number density of LGRBs between the redshifts $z_{1}$ and $z_{2}$~(see \citealt{Elliott12a} and references therein), and $M_{\mathrm{min}}$ and $M_{\mathrm{max}}$ are the minimum and maximum star masses considered in our model respectively. We note that corrections based on instrument limitations (e.g., LGRB luminosity function, jet opening angles, and redshift completeness) are not dealt with at this step and left to be deconvolved from the sample used.

\subsubsection{Biasing Proxies}
\label{sec:ModelandMethodology:subsec:ProxyModel:subsubsec:BiasingProxies}
The majority of LGRB metallicity observations usually probe the global physical properties of the host galaxy, rather than the direct LGRB progenitor environment, mainly due to the signal-to-noise of emission lines being stronger for the entire galaxy than the LGRB location, such that high-spatial resolution cannot be achieved~\citep[see e.g.,][for a situation contrary to this]{Christensen08a,Thoene14a}. As a result, the constraints of the LGRB progenitor model (i.e., $Z_{\rm progenitor}<0.3\, \rm Z_{\odot}$ or those outlined in Sect.~\ref{sec:Methodology}) are commonly used on the host galaxy's physical properties via some analytical proxy to correct for any bias when calculating the LGRB rate from the CSFH. This technique is commonly used to estimate any environmental bias between the two rates~\citep[see, e.g.,][]{Elliott12a}, i.e., a metallicity cut-off in host galaxy's~\citep[see, e.g.,][]{Salvaterra12a}. However, it is possible that the quantities derived from such studies do not reflect the physical nature of the environment of the LGRB nor the host galaxy it prefers. To this end, we investigate a handful of proxies used on LGRB host galaxies commonly used in the literature. We then compare the calculated LGRB rates to those determined from our physically motivated model, and also using the same physical constraints on the star particles within the simulation. The following proxies are investigated:
\begin{enumerate}[i)]
  \item {\bf BP1}: a metallicity cut-off such that galaxies, at a given redshift $z$, with a gas-phase metallicity above $Z_{\rm gal,cut}$ do not produce LGRBs. When convolved with a galaxy luminosity function, the fraction of the CSFH that contributes to the LGRB rate, with a cut of $Z_{\rm gal,cut}$, is described by the following {\bf analytical formula}~\citep{Langer06a}:

    \begin{eqnarray}
      \Psi(z, Z_{\rm gal, cut}) = \frac{\hat{\Gamma}\left[0.84, \left(Z_{\rm gal, cut}/\mathrm{Z}_{\odot}\right)^{2} 10^{0.3z}\right]}{\Gamma(0.84)}
    \end{eqnarray}

    \noindent where $\Gamma$ and $\hat{\Gamma}$ are the gamma and incomplete gamma functions, respectively. Note, that this is folded directly with the CSFH.

  \item {\bf BP2}: a strict metallicity cut-off, such that the CSFH is only calculated from the {\bf star particles} within the {\bf simulation} that exist in a host galaxy with a gas-phase metallicity below $Z_{\rm gal,cut}$~\citep[e.g.,][]{Campisi11b}. Note, that this directly interacts with the star particles of the simulation, unlike BP1.
	\item {\bf BP3}: a {\bf probabilistic} metallicity bias, such that the more metal-rich galaxies contribute fewer LGRBs than low-metal galaxies, written mathematically as~\citep{Trenti13a}:

    \begin{eqnarray}
      \Psi(Z_{\rm gal})=\frac{a\log_{10}\left(Z_{\rm gal}/\rm Z_{\odot}\right)+b+0.3}{1.3}\\
			\left[a,b\right] = \left\{\begin{array}{rl}
			\left[0,1\right]&{\rm if}\quad Z/\rm Z_{\odot}\leq10^{-3}\\
			\left[-\frac{3}{8},-\frac{1}{8}\right]&{\rm if}\quad 10^{-3}<Z/\rm Z_{\odot}\leq10^{-1}\\
			\left[-\frac{1}{4,0}\right]&{\rm if}\quad 10^{-1}<Z/\rm Z_{\odot}\leq1\\
			\left[0,0\right]&{\rm if}\quad Z/\rm Z_{\odot}>1\\
			\end{array}\right. \nonumber
    \end{eqnarray}

\noindent Note, that this interacts with the {\bf subhalos} of the simulation to recalculate the CSFH.
  
	\item {\bf BP4}: only {\bf star particles} that have a metallicity below $Z_{\rm cut}$ are selected and contribute to the determination of the CSFH, and thus the LGRB rate. Note, that this interacts with {\bf star particles} directly, like BP2.
\end{enumerate}

For each of the four proxies, we first determine the CSFH from the simulation (folded with an above bias if required) and then transform this into a LGRB rate with Eqn.~\ref{eqn:csfh_lgrbr}, including the respective bias. We assume a value of $Z_{\rm gal, cut}=Z_{\rm cut}=0.1\, \rm Z_{\odot}$ for the biases that require a fixed value, which was selected to remain consistent with both cosmological and proxy studies.


\section{Results}
\label{sec:Results}

\subsection{LGRB Rates Within the Simulation and Comparison With Observations}
\label{sec:Results:SimulatedLGRBR}

\begin{figure}
  \begin{center}
    \includegraphics[width=9.5cm]{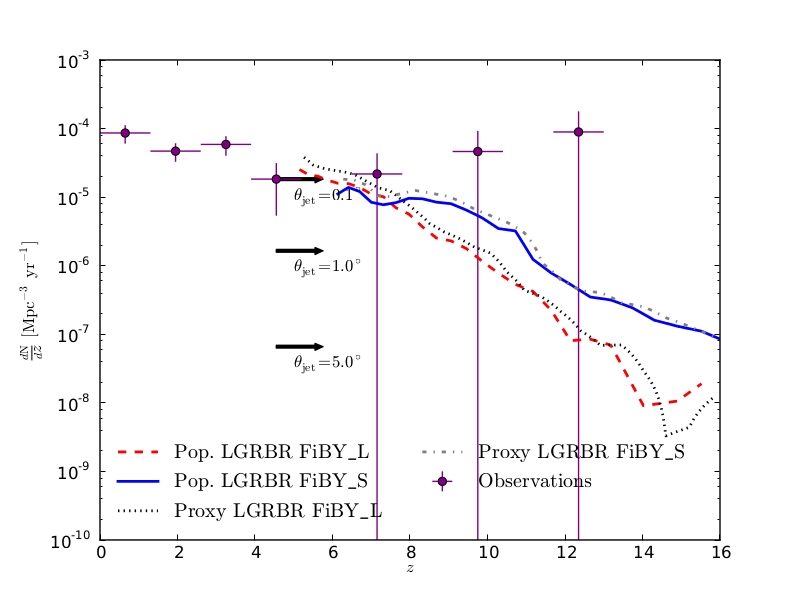}
    \caption{Number density of LGRBs per redshift bin determined from the progenitor model (FiBY\_L, red-dashed line; FiBY\_S blue line), the proxy model (FiBY\_L, black-dotted line; FiBY\_S grey-dash-dotted line), and observations obtained by GROND (purple dots; see Sect.~\ref{sec:Results:SimulatedLGRBR}). The black arrows show the change in the normalisation to the observed data points for the adopted LGRB jet opening angle~\citep[$\theta_{\rm jet}=0.1-10\degr$, e.g.,][]{Yonetoku05a}.}
    \label{fig:lgrbr_fiby_csfh_obs}
  \end{center}
\end{figure}

The number density of LGRBs from both the progenitor and (unbiased) proxy model can be seen in Fig.~\ref{fig:lgrbr_fiby_csfh_obs}. It is clearly seen that the proxy LGRB rate reproduces well the trend of the progenitor LGRB rate despite the mass-metallicity selection constraints placed on the progenitor LGRB rate. The deviation between the two simulation boxes is merely a result of resolution and does not affect the conclusions. Already, this demonstrates that under the progenitor framework used the LGRB rate would be a good tracer of the CSFH at redshifts of at least $z>5$. This is consistent with recent work in the radio and infra-red regimes that show that the LGRB rate does, in fact, trace the CSFH in an unbiased way for redshifts of $0<z<2.5$~\citep{Michalowski12a, Hunt14a, Perley14a}. If no bias is present, this would make it very simple to convert the LGRB rate to the CSFH, without having to be concerned with the underlying progenitor model. A first approximation would only require the mass cut-off for creating a black hole, which could be constrained by, e.g., mass yields determined from LGRB supernovae~\citep[see, e.g.,][]{Olivares12a}.

We use the LGRB sample compiled with LGRBs that were observed with the Gamma-ray Burst Optical/Near-infrared Detector (GROND) within 4 hours of their initial detection~\citep{Greiner11a}, to compare to the progenitor LGRB rate of the simulation. The GROND upper limit 2 sample (outlined in~\citealt{Elliott12a}) has a redshift completeness of 95\% and has a redshift range $z=0-12$. The number density is calculated by taking the histogram of the sample and dividing each bin by its size and redshift volume. The absolute rate is then calculated by deconvolving the distribution by: (i) the observation length, $\Delta T=3.5\,\rm yr$, (ii) the fraction of the observable sky, $\frac{\Omega_{\rm GROND}}{4\pi}=0.077$, (iii) the LGRB luminosity function, and (iv) the beaming correction, $\eta_{\rm jet}=1.5\times10^{-4}$. For (iii) we utilise the best-fit luminosity function in~\citealt{Elliott12a} and deconvolve for each individual redshift bin, including the luminosity limit of GROND. We adopt the best-fit opening angle for (iv) of $\theta=1\degr$. However, we note that the accepted range of opening angles varies between $0.1-10\degr$~\citep[e.g.,][]{Yonetoku05a}, for which we show its affect on the underlying observed LGRB rate in Fig.~\ref{fig:lgrbr_fiby_csfh_obs}.

The progenitor- and proxy-model determined LGRB rate show a preference for small opening angles of the order or $\theta=0.1\degr$ and would suggest the LGRB rate traces that of the CSFH (for $z>5$). However, if there was a stronger metallicity selection on the LGRB progenitor, the LGRB rate would prefer larger opening angles (this degeneracy is discussed in more detail in Sect.~\ref{sec:Discussion:ModelLimitations}).

\subsection{Effects of Biasing Proxies}

\begin{figure*}
  \begin{center}
    \includegraphics[width=9.5cm]{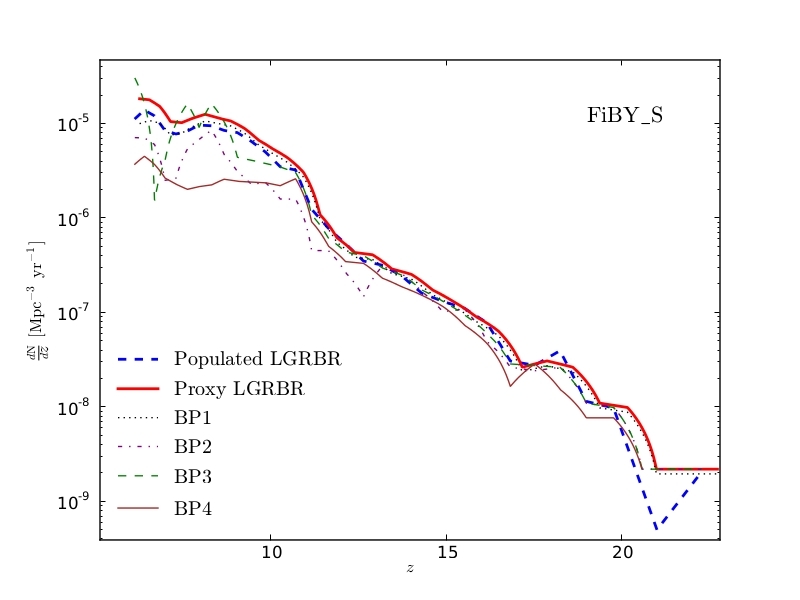}\includegraphics[width=9.5cm]{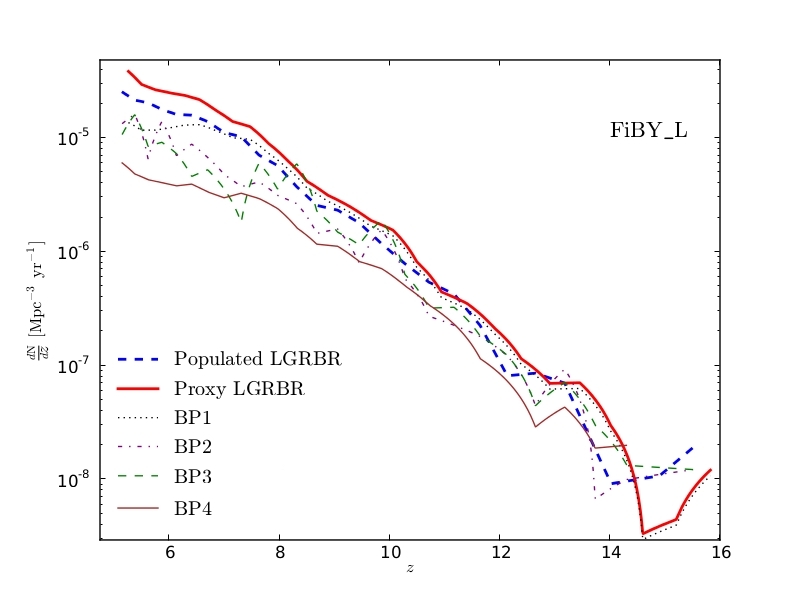}
    \caption{Evolution of the LGRB rate across redshift for both the {\it progenitor} and {\it proxy} models (see Sect.~\ref{sec:Methodology}), and also for both of the boxes investigated $4^3\, \rm Mpc^3$ (FiBY\_S, left) and $16^3\, \rm Mpc^3$ (FiBY\_L, right). For comparison the LGRB rate determined for each of the biasing proxies considered are also displayed. Their definitions can be found in Sect.~\ref{sec:ModelandMethodology:subsec:ProxyModel:subsubsec:BiasingProxies}.}
    \label{fig:lgrbr_proxy_comparison}
  \end{center}
\end{figure*}

We implement the four biasing proxies outlined in Sect.~\ref{sec:ModelandMethodology:subsec:ProxyModel:subsubsec:BiasingProxies} to the LGRB rate discussed in the previous section. Figure~\ref{fig:lgrbr_proxy_comparison} shows the four resulting rates determined from BP1, BP2, BP3, and BP4.

BP1 begins to deviate from the progenitor LGRB rate at a redshift of $z\sim8$. This is expected, as for redshifts of $z>8$ the analytical expression is unity, such that all galaxies contribute to the CSFH (or the empirical fits have no galaxies above $Z_{\rm gal,cut}$, above this redshift). An extrapolation from the last redshift simulation-output ($z\sim5$) of BP1 would suggest that it would continue to move away from the progenitor LGRB rate.

BP2 and BP3 have similar behaviour to BP1, except that the simulation begins to form host galaxies of $Z>Z_{\rm gal,cut}$ much earlier (as compared to those from the empirical function of BP1), at redshifts of $z\sim14$. The LGRB rate of BP2 and BP3 then begin to deviate away from the progenitor LGRB, reaching similar values of BP1 at the last redshift simulation-output ($z\sim6$).

Finally, BP4 systematically underestimates the progenitor LGRB rate from the first redshift simulation-output ($z\sim16$) and gradually deviates away up until the final redshift simulation-output ($z\sim6$).

Most importantly, all of the BPs differ from BP4 and also show that they would continue to deviate even further as approaching $z=0$. BP4 directly selects star particles below a given metallicity cut-off (at a mass resolution of $\sim10^{3}\, \rm M_{\odot}$) and should be reflected by any of the three proxies, BP1, BP2, or BP3, which clearly is not the case. This result already shows the proxies are not directly related to the environmental properties of the LGRB, but is in fact caused by different physical properties of the host (we discuss this further in Sect.~\ref{sec:Discussion:MetalRichProg}).

\subsection{LGRB Host Galaxies and Progenitor Environments}

\begin{figure*}
  \begin{center}
    \includegraphics[width=10.cm,clip=True,trim=0cm 0cm 0cm 0cm]{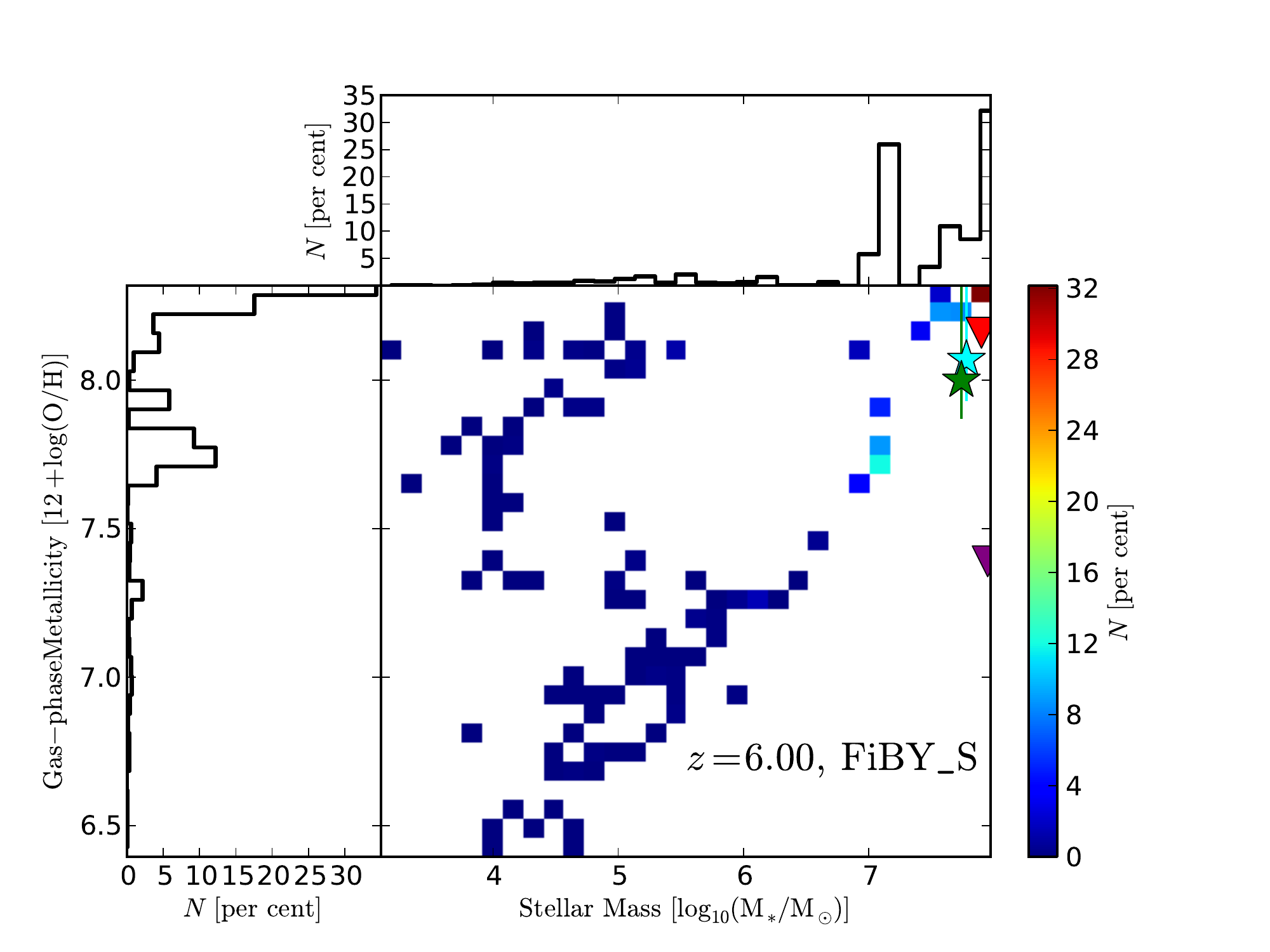}\includegraphics[width=10.cm,clip=True,trim=0cm 0cm 0cm 0cm]{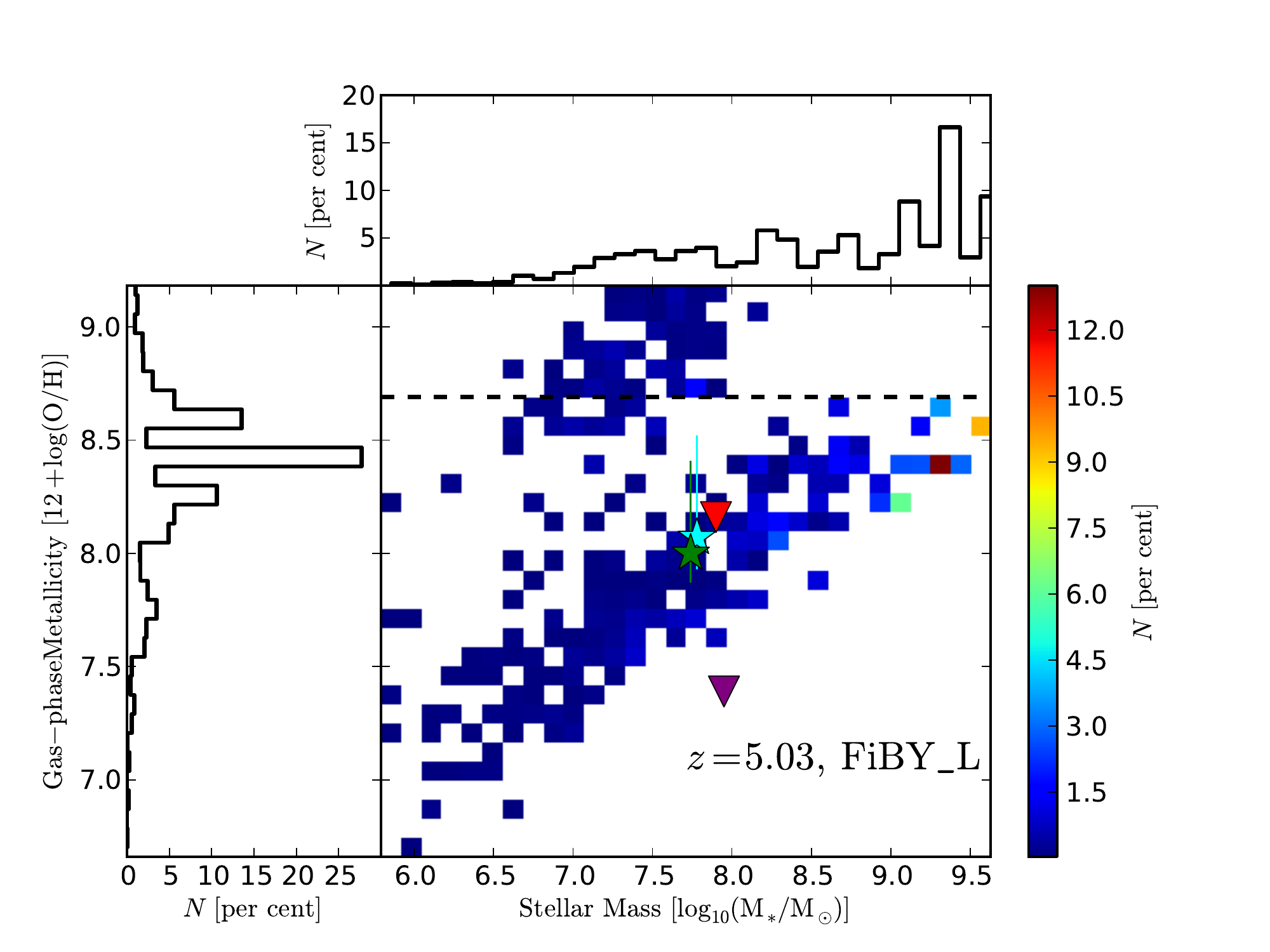}
    \caption{Gas-phase metallicity vs. stellar mass for each host associated to a LGRB. The colour bar denotes the fractional number of LGRBs per each type of galaxy. The histograms denote the percentage of hosts that contain a given amount of metals or stellar mass. The horizontally dashed black line depicts a metallicity of $1\, \rm Z_{\odot}$, assuming solar values of~\citealt{Asplund09a}. The four coloured marks denote four observed LGRB host galaxies: GRB 130702A~\citep[red down triangle;][]{Kelly13a}, GRB 060614~\citep[purple down triangle;][]{DellaValle06a}, XRF060218~\citep[cyan star;][]{Pian06a, Levesque10d}, and GRB 030329~\citep[green star;][]{Levesque10d}.}
    \label{fig:mass_metallicity_OH_2D}
  \end{center}
\end{figure*}

We determine the LGRB host galaxy associated with the underlying progenitor, by finding the gravitationally bound subhalo the star particle exists in. The physical properties of the host are then determined by all the star particles that are associated to this subhalo. We plot the gas-phase metallicity and stellar mass of each galaxy hosting a viable LGRB in Fig.~\ref{fig:mass_metallicity_OH_2D}. Two distinct areas of the phase-space are easily seen. First, the standard mass-metallicity relationship, and second, a collection of low-mass ($10^{6.5-8.0}\, \rm M_{\odot}$), high-metallicity ($Z>1.0\, \rm Z_{\odot}$) satellite galaxies.

\begin{figure*}
  \begin{center}
    \includegraphics[width=9.cm]{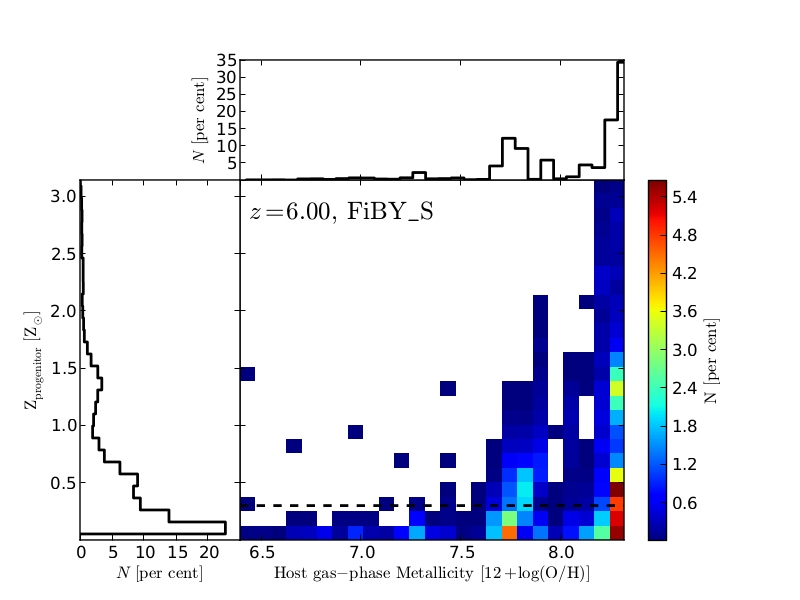}\includegraphics[width=9.cm,clip=True,trim=0cm 0cm 0cm 0cm]{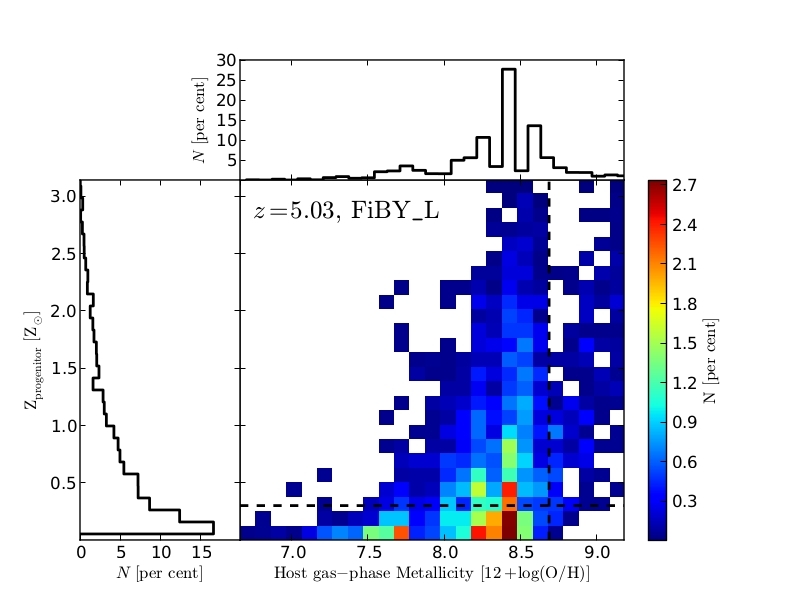}
    \caption{Progenitor metallicity of the LGRB plotted against the gas-phase metallicity of its host galaxy. The colour bar denotes the percentage of LGRBs and hosts that have these quantities. The histograms depict the independent percentages of progenitor metallicity and host galaxy metallicity. The vertical dashed line is a metallicity of $1\, \rm Z_{\odot}$, assuming solar values of~\citealt{Asplund09a} and the horizontal line depicts a metallicity of $0.3\, \rm Z_{\odot}$, which is the progenitor limit expected from collapsar models~\citep[see, e.g.,][]{Hirschi05a}.}
    \label{fig:prog_metal_gal_metal_2D}
  \end{center}
\end{figure*}

Secondly, we are interested in how the progenitors environments vary with, or depend on, the global properties of the host galaxy. Therefore, we plot the stellar metallicity of the LGRB to the corresponding gas-phase metallicity of the host galaxy that it resides in, in Fig.~\ref{fig:prog_metal_gal_metal_2D}. In FiBY\_L it is seen that the LGRB progenitor metallicities exist in a wide range of host galaxy metal environments, with a large fraction that have a gas-phase host metallicity of $\sim12+\log\left(\rm O/H\right)\sim8.5$. This is also confirmed in the FiBY\_S simulation box. Most interestingly, if we culled any progenitor that had a metallicity of $Z<0.3\, \rm Z_{\odot}$ (horizontal black dashed line in Fig.~\ref{fig:prog_metal_gal_metal_2D}) we would still have the full range of host galaxy metals. This is an underlying property of the metal distribution of the host galaxies, which we discuss in more detail in Sect.~\ref{sec:Discussion:MetalRichProg}.

\section{Discussion}
\label{sec:Discussion}

\subsection{Model Limitations}
\label{sec:Discussion:ModelLimitations}
Each assumption made in our progenitor model has an underlying effect on the overall quantity and rate of LGRBs that are formed. Firstly, we adopted a non-evolving stellar IMF of the Salpeter form. Given the static nature of alternative choices of IMFs, any other type would translate to a constant offset in the absolute LGRB rate. Such a change would not alter the evolution of the LGRB rate and would remain in the uncertainties of the other parameters involved, such as the LGRB opening jet angle. For example, changing to a Chabrier IMF would alter the rates by $\sim0.2\, \rm dex$, in contrast to the opening jet angle that easily changes by $\sim3\, \rm dex$. The use of an evolving IMF with redshift would change the results drastically. For example, assuming a double power law IMF with a break that evolves with redshift as $M_{\rm break}\propto(1+z)^2$ would mean that there would be a factor 5 more $40\, \rm M_{\odot}$ stars produced in a gas cloud at a redshift of $z=10$ compared to a redshift of $z=3$~\citep[see, e.g.,][]{Dave08a}. This would result in a larger number density of LGRBs at higher redshifts in our current framework.

The contribution of the jet opening angle of the LGRB emission varies by three orders of magnitude (i.e., $0-10\degr$) as already seen in Fig.~\ref{fig:lgrbr_fiby_csfh_obs}, given the wide range of measurements determined from slope changes in LGRB afterglow light curves~\cite[see, e.g.,][]{Racusin09a}. Unfortunately, recent simulations have shown that this change in the afterglow light curve may occur at much later times than currently thought, and therefore, the change in the light curve that we currently observe is not in fact due the result of the jet opening angle~\citep{vanEerten12a}. As a result, there are currently no robust measurements of the opening angle. This means there will still remain a degeneracy between the opening angle and any underlying bias between the CSFH and the LGRB rate. However, this degeneracy is still dominated by the adopted value of the opening angle. Without a precise value for the opening angle, any difference in progenitor model would be indistinguishable when investigating the LGRB rate.

\begin{figure}
  \begin{center}
    \includegraphics[width=9.5cm]{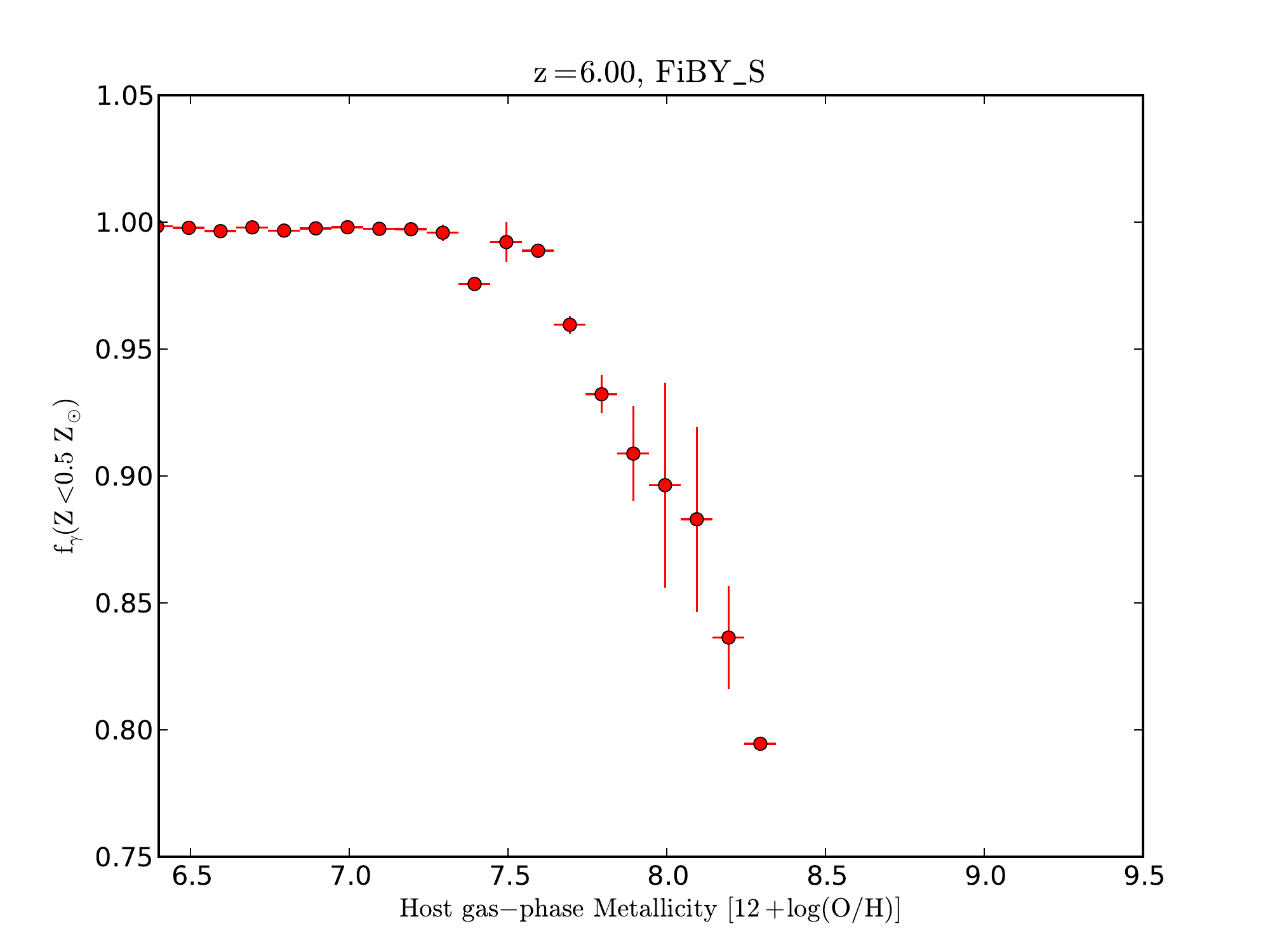} \includegraphics[width=9.5cm]{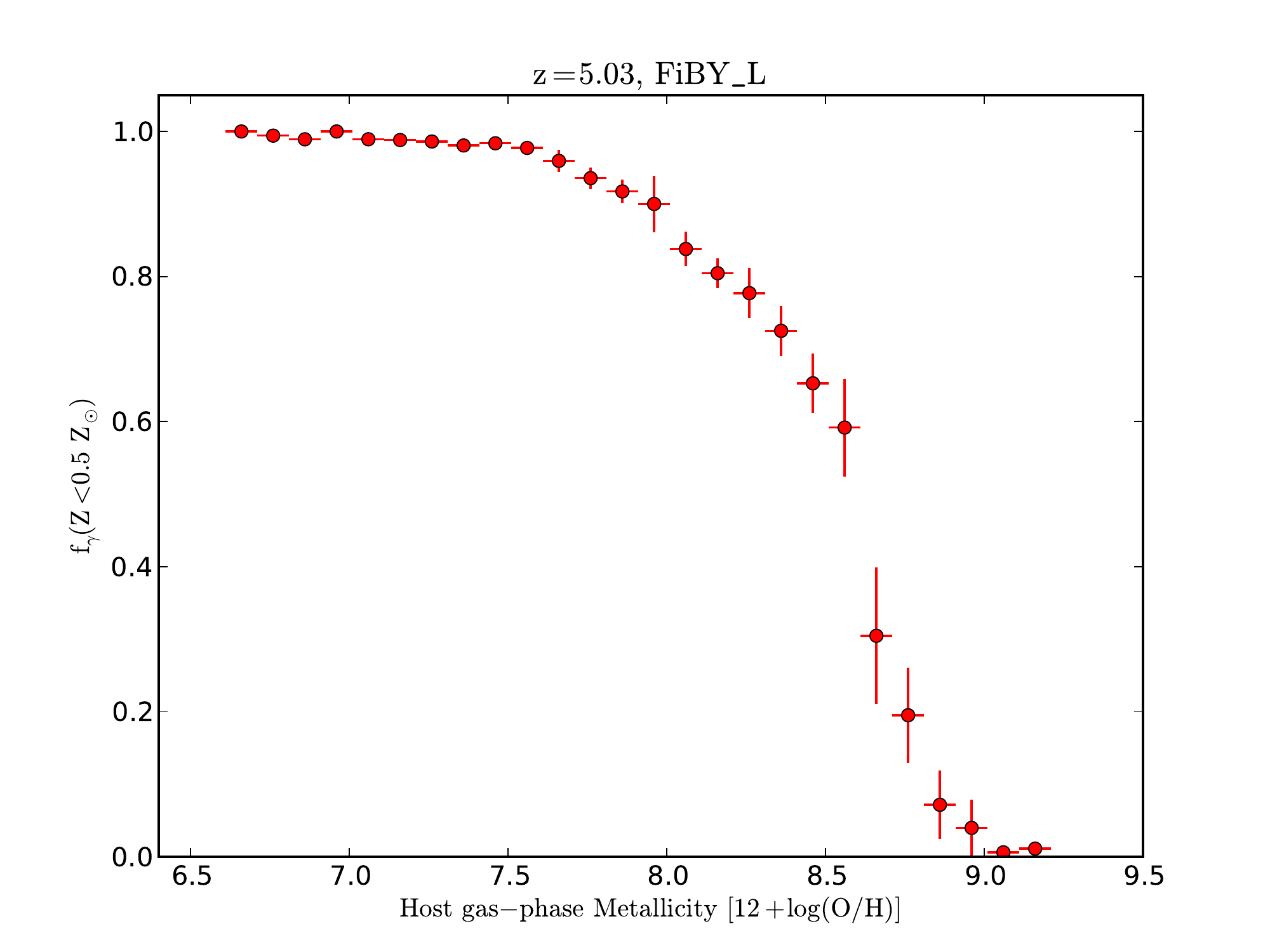}
    \caption{Fraction of gas mass that a galaxy has below a given threshold value $Z_{\rm gal,cut}$ vs. its integrated gas-phase metallicity. We note that the top panel is a zoom-in and there exists no fractional values below $\sim0.8$.}
    \label{fig:f_gamma_1D}
  \end{center}
\end{figure}

In addition, the mechanism by which a LGRB will form is still a lively debated topic and would also have an effect on our conclusions. Currently, the collapsar model is the most favoured mechanism, for which we still do not know the exact mass (e.g., $M_{\rm BH}$) or metallicity requirements to result in the formation of a LGRB. Also, recently it has been shown that almost $50\%$ of stars are in binaries~\citep[e.g.,][]{Sana13a}. The possibility of a LGRB forming in a binary would completely change the requirements of the LGRB formation mechanism, and would not be entirely dependent on the metallicity of the environment~\citep[e.g.,][]{Fryer99a}. Changing the progenitor model, especially to one that depends on parameters that are evolving through redshift, would alter the results found in this paper. To this end we also consider a simple metallicity cut, by which, we further select LGRBs as those that occur in regions of purely low metallicity, i.e., with a metallicity of $Z<\rm 0.3\, Z_{\odot}$. As seen in both Figs.~\ref{fig:prog_metal_gal_metal_2D}, we see that regardless of this metallicity selection we would still have a broad range of host galaxy metallicities, and would not exclude the metal-rich hosts of $Z\sim1\, \rm Z_{\odot}$. As already noted, this is a result of the distribution of star-forming metal-poor gas within each host galaxy and so we plot the fraction of the gas in a host galaxy that has a metallicity below $Z<\rm 0.5\, Z_{\odot}$ in Fig.~\ref{fig:f_gamma_1D}. Even galaxies that are highly enriched have a large fraction ($10-50\%$) of metal poor environments to allow a LGRB to form in a simple threshold model, showing our results are unchanged by adopting a more rigid progenitor model.

Finally, we note that the collapsar model is not the only avenue of forming a LGRB. Other possibilities exist that involve: (i) convective mixing of the outer layers~\citep{Frey13a}, (ii) binary systems~\citep[e.g.,][]{Fryer99a}, and (iii) magnetars~\citep{Thompson95a}. Incorporation of such progenitor models is possible in our studies, but deeper understanding of the progenitor mechanism is required in order to parametrise the evolutionary effects. Obtaining detailed information of the local LGRB's environment is already a difficult task, especially at redshifts of $z>5$ and so other indirect methods of determining the progenitors environments can help shed light on the mechanism, for which we discuss more in Sect.~\ref{sec:Discussion:MetalRichProg}.

\subsection{Simulation Caveats}
\label{sec:Discussion:SimulationCaveats}
Our results depend upon the physical quantities derived within the simulation, specifically the metallicity. The enrichment of metals within the simulation is dominated by the feedback from the stellar winds and supernovae explosions of both the Population-III (pair-instability supernovae) and -II (supernovae type-II) stars. Metal yields of the supernovae are determined from tabulated datasets based on stellar evolution calculations~\citep{Portinari98a, Marigo01a, Heger02a, Heger10a}. Given the complexities involved, e.g., convective boundaries, mass loss, dredge-up, and hot-bottom burning it is possible that the quantity of metals could be larger or lower than that determined in both of the simulation boxes. Such effects would change the results by a factor of $\sim2$~\citep{Wiersma09a}, which in terms of gas-phase metallicity, would not alter our underlying results. More importantly, an alternate IMF would also be very sensitive for the low-mass slope, as this would result in a higher or lower quantity of metal pollution from AGB stars and stellar winds, modifying the amount of carbon and nitrogen. However, such IMFs are not directly observed~\citep[][]{Bastian10a} and the quantity of metals would not be changed for IMFs similar to a Chabrier or Salpeter type~\citep{Wiersma09a}. Also, it can be seen from Fig.~\ref{fig:mass_metallicity_OH_2D} that both simulation boxes show similar mass-metallicity trends and the clustering of the low-mass, high-metallicity galaxies, for their respective redshifts. 

\begin{figure}
  \begin{center}
    \includegraphics[width=9.cm]{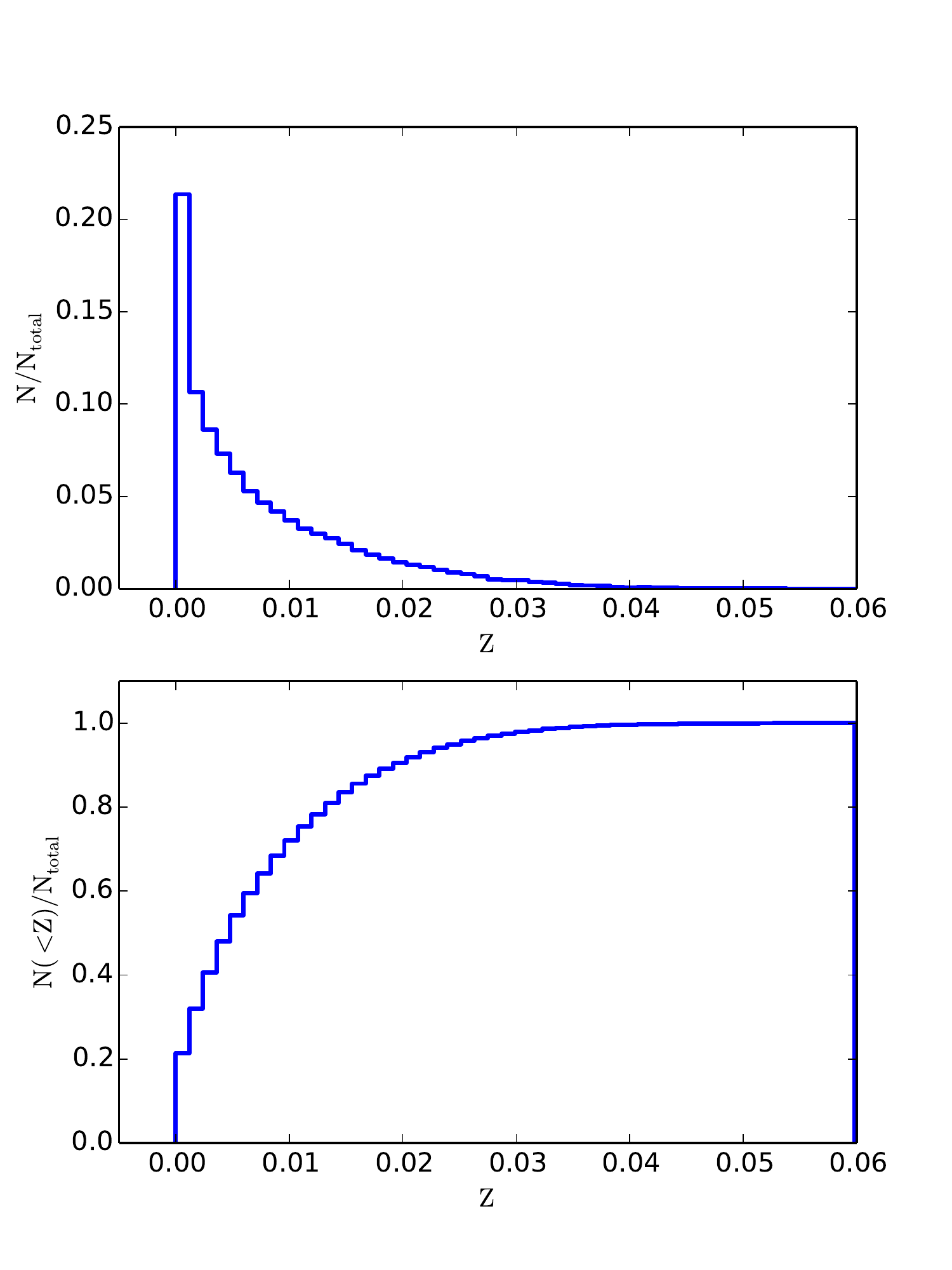}
    \caption{{\bf Top}: Histogram of stellar metallicities of star particles within the FiBY\_L simulation at a redshift of $z=5.03$. {\bf Bottom}: Cumulative distribution of stellar metallicities of star particles within the FiBY\_L simulation at a redshift of $z=5.03$.}
    \label{fig:metal_dist}
  \end{center}
\end{figure}

Finally, we note that the progenitor model depends heavily upon the redshift at which large enough metals can be produced, such that the metallicity cuts shown in Fig.~\ref{fig:mass-metallicity} have any effect. Metals of $Z\sim0.01$ start being produced at around $z\sim8.5$ and so, inherently, the progenitor model will predict that LGRBs will trace the CSFH at redshifts larger than this as opposed to the collapsar model. We show the distribution of stellar metallicities within the simulation in Fig.~\ref{fig:metal_dist}, which demonstrates that there are LGRBs with large enough metals to be affected by the progenitor model selection.

\subsection{Metal Rich Progenitors in Satellite Galaxies}
\label{sec:Discussion:MetalRichProg}
The standard mass-metallicity relation of galaxies can be seen in Fig.~\ref{fig:mass_metallicity_OH_2D} for both the FiBY\_L and FiBY\_S simulation boxes (see also Dalla Vecchia et al. in preparation). The dispersion of metals within these host galaxies, as depicted in Fig.~\ref{fig:f_gamma_1D}, means that the metallicity of the LGRB progenitor sites cover a wide range of values as seen in Fig.~\ref{fig:prog_metal_gal_metal_2D}, i.e., an environment with a metallicity of $\sim0.3\, \rm Z_{\odot}$ can exist in any host with a range of global metallicities of $12+\log\left(\rm O/H\right)\sim7$ to $12+\log\left(\rm O/H\right)\sim9$. Unfortunately, this means that if a LGRB is observed in a massive, metal-rich galaxy that sits on the mass-metallicity relation, it would not be certain if the environmental metallicity is high or low without metallicity measurements obtained directly at the site of the LGRB. Such measurements become difficult for high redshift ($z>1$) events due to the signal-to-noise of the emission lines at the location of the LGRB, which are usually offset from the galaxy centroid by a median value of $\sim1\, \rm kpc$~\citep{Bloom01a}. Therefore, we note caution on results that are drawn on progenitor models based on the host galaxy's global properties. 

We note a class of metal-rich and mass-poor galaxies seen in Fig.~\ref{fig:mass_metallicity_OH_2D} that lie above $12+\log\left(\rm O/H\right)\sim8.7$ and below $M_{*}\sim10^{8}\, \rm M_{\odot}$. The LGRB environments of these hosts lie in a unique position of progenitor environment vs. host galaxy environment phase space (Fig.~\ref{fig:prog_metal_gal_metal_2D}), such that they have no low-metal progenitor environments despite there existing metallicity dispersions within the host. A discovery of such a type of LGRB host galaxy would give direct evidence that a LGRB occurred in a metal-rich environment, which would oppose the metallicity requirements commonly placed on the collapsar model. Currently, no host with such properties has been observed, even at low redshifts, and there are no secure detections of LGRB host galaxies for $z>5$~\citep{Tanvir12a}. Therefore, only comparisons with low redshifts can be carried out, with host galaxies that resemble some of the properties of those predicted. To this end, we collect four of the lowest mass LGRB host galaxies ($M_{*} < 10^{8}\, \rm M_{\odot}$.): GRB 130702A~\citep{Kelly13a}, GRB 060614~\citep{DellaValle06a}, XRF060218~\citep{Pian06a, Levesque10d}, and GRB 030329~\citep{Levesque10d}, and plot them in Fig.~\ref{fig:mass_metallicity_OH_2D}. They are seen to lie on the mass-metallicity relation defined by the simulation, but are still offset from the galaxies of larger metallicities. 

The high-metallicity population exhibits SFRs of $0.1$ to $1.4\, \rm M_{\odot}/yr$. It is most likely these are post-starburst galaxies left with little gas and are, therefore, heavily enriched by metals. As is already seen in several models of the galaxy stellar mass-metallicity relation, their metallicity will decrease when accreting new gas~\citep[for a review, see][]{SanchezAlmeida14a}. Even if the time scale over which they can be observed might be short, depending on the next accretion event, they should still exist for some time. Currently, there are no observations of metallicities in low mass dwarfs at high redshift for comparison. However, low redshift observations demonstrate, independent of the formation mechanism, that such type of galaxies are plausible to exist~\cite[see, e.g.,][]{Sweet14a}. Finally, a post-starburst galaxy at $z=0.211$, with little or no star formation, has found to be a LGRB host galaxy~\citep[GRB 050219A;][]{Rossi14a}.

The galaxies investigated in our studies lie at redshifts of $z\sim5$, which makes the standard techniques used to determine metallicities, such as the detection of emission lines, e.g., [OII] ($\lambda\lambda 3726,3729$), [NII] ($\lambda 6584$) and [OII] ($\lambda\lambda 4959,5007$), more difficult as they are redshifted to wavelengths of $\sim2-4\, \rm \mu m$. At such wavelengths, infra-red spectrographs are required. Detection of these lines are possible with instruments currently mounted at the VLT, e.g., CRIRES\footnote{ \url{www.eso.org/observing/etc/bin/simu/crires} }. However, to acquire a signal-to-noise greater than $3$ for an emission line of $f_{\rm [OII]}\sim10^{-17}\, \rm erg\, s^{-1}cm^{-2}$ at a wavelength of $3.72\, \mu \rm m$ would require exposure times of the order of $100$ hours. In contrast, when the {\it James Webb Space Telescope} is launched, it will contain an infra-red spectrograph, NIRSpec{\footnote{\url{jwstetc.stsci.edu/etc/input/nirspec/spectroscopic}}}, and would be able to easily get a signal-to-noise of $25$ for the same source on the order of tens of minutes. In addition, stellar masses would also need to be determined, which, although challenging, could be determined via standard spectral energy distribution techniques using photometry~\citep[see, e.g.,][]{Bruzual03a}. Utilising the hosts obtained from our population model, we would expect that $10\%$ of these type of galaxies would exist at a redshift of $z>5$. Given the current rate of LGRBs at these redshifts of $\sim1\, \rm yr^{-1}$, would suggest that we would find $0.1\, \rm yr^{-1}$. For the $\sim10$ years that {\it Swift} has observed would imply that we should have at least one LGRB host of this kind for this redshift. However, we currently only have upper limits for high redshift host galaxies~\citep{Tanvir12a}.


\section{Conclusion}
\label{sec:Conclusion}

We utilised two cosmological simulations with a wide range of resolutions from the First Billion Years simulation project. LGRBs were populated in the simulations (post-hoc) utilising a combination of Monte-Carlo techniques and a physically motivated progenitor model that depends on micro-parameters of its environment rather than the global properties of the host galaxy. We investigated the resulting LGRB rate and its comparison to the CSFH, and also the global and local environments of the newly formed LGRB. The following main conclusions are drawn:

\begin{enumerate}[1.]
  \item Using a physically motivated LGRB progenitor model, we show that the LGRB traces the CSFH at redshifts of $z>5$.
  \item Analytical proxies that try to match the CSFH to the LGRB rate based on global host properties of the LGRB host galaxy, do not in fact reflect the local physical properties of the underlying progenitor environment.
  \item Even at redshifts of $z=5$ we expect to see metal-rich LGRB host galaxies.
  \item Detection of a metal-rich, low-mass satellite host galaxy would directly show a LGRB that has formed in a metal-rich environment and directly argue against the collapsar model.
\end{enumerate}


\section*{Acknowledgements}
We thank the referee for their constructive and helpful suggestions. JE thanks A. Longobardi, B. Agarwal, J.~-P. Paardekooper, D. A. Kann, H. van Eerten, and D. Burlon for their help. CDV acknowledges support by Marie Curie Reintegration Grant PERG06-GA-2009-256573.

\label{lastpage}

\bibliographystyle{mn2e}
\bibliography{main}

\end{document}